\documentclass[twocolumn, showpacs,preprintnumbers,amsmath,amssymb,pra]{revtex4}
\usepackage{amssymb}
\usepackage{amsfonts}
\usepackage{graphicx}
\usepackage{dcolumn}
\usepackage{bm}

\usepackage{times,epsfig,amssymb,amsmath}

\begin{document}

\title{Thermalization from a general canonical principle}

\author{Shuai Cui$^{1,2}$, Jun-Peng Cao$^2$, Hui Jing$^1$,
Heng Fan$^2$\footnote{hfan@iphy.ac.cn} and Wu-Ming Liu$^2$}%
\affiliation{%
$^1$Department of Physics, Henan Normal University, Xinxiang 453007, China\\
$^2$Beijing National Laboratory for Condensed Matter Physics,
Institute of Physics, Chinese Academy of Sciences, Beijing
100190, China
}

\date{\today}

\begin{abstract}
We investigate the time evolution of a generic and finite isolated quantum many-body system
starting from a pure quantum state.
We find the kinematical general canonical principle
proposed by Popescu-Short-Winter for statistical mechanics can be built in
a more solid ground by studying the thermalization, i.e.
comparing the density matrices themselves rather than the measures of distances.
In particular, this allows us to explicitly identify that,
from any instantaneous pure state after thermalization, the state of
subsystem is like from a microcanonical ensemble
or a generalized Gibbs ensemble, but neither a canonical nor a thermal ones due to finite-size effect.
Our results are expected to bring the task of characterizing the state
after thermalization to completion. In addition,
thermalization of coupled systems with different temperatures
corresponding to mixed initial states
is studied.
\end{abstract}
\maketitle

\section{Introduction}
Recently, studies of the foundation of
statistical mechanics
are enjoying a renaissance.
Concerning about the canonical ensemble, it was pointed
out that the equal a priori probability postulate of statistical mechanics \cite{Landau},
which is applied to the microcanonical ensemble constituted by
all pure states satisfying the energy constraint, is not necessary.
Instead, it should be replaced by the general canonical principle \cite{winter}
or canonical typicality \cite{Goldstein}. It states that
the state of a subsystem
can be obtained from a single pure state rather than a mixture of
microcanonical ensemble. Explicitly, suppose we have a weakly coupled
system constituting by system `S' and the heat bath `B',
in case that the heat bath `B' is large
enough, starting from
almost every pure state of the coupled system `SB', the state of the system
`S' is approximately the canonical density matrix.
The interest in these fundamental questions is also
due to the fact that those postulates can be directly
realized in recent experiments of ultracold quantum gases
\cite{Blochreview,Kinoshita,Hofferberth}.
Some related conclusions are obtained from different points of view
\cite{RigolNature,Srednicki,Tasaki,sun,Rigol2007,Altman,Noack,Lee,
wigner,cirac,Rigol2011,plenio,BKL,Jens,book}.

The general canonical principle is of kinematic rather than dynamic \cite{winter,Goldstein} since
it does not involve any time evolution of the state. Also the canonical density matrix
from a pure state is because of the entanglement between system and the heat bath,
otherwise the system will also
be a pure state which cannot equal to a canonical ensemble.
Dynamically, to result in a pure state between the coupled `SB' system,
the system as a whole should be an isolated
system and also the initial state should be pure. And more importantly, all
processes are of quantum mechanics such as the involvement of entanglement and
Schr\"odinger equation for time evolution. Recently remarkable progress has
been made in understanding the temporal evolution of the isolated many-body
quantum system. In Ref.\cite{RigolNature}, Rigol \emph{et al.}
show that in a generic isolated system, non-equilibrium dynamics is
expected to result in thermalization: any pure states will
relax to pure states in which the values of macroscopic quantities are stationary
and universal. In particular, the thermalization can happen at the level of individual
eigenstates. However, the characteristic of the state after thermalization
seems not clear, in particular about the relationship
between the general canonical principle and the process of thermalization.

In this paper, we will study a generic isolated quantum
many-body system from the view points of both kinematics and dynamics.
By using a generic thermalization model, we prove the general
canonical principle numerically by showing that
the state of `S' from instantaneous pure state after thermalization
is similar as from a microcanonical
ensemble.
Our main method is based on comparing the exact form of density matrices,
rather than the measure of distance or the one-dimensional marginal
momentum distribution, thus the results are more
explicit and more precise. We then can clarify a crucial question that the state of
a subsystem in dynamical equilibrium is like from a microcanonical ensemble
or a generalized Gibbs ensemble, but neither a canonical nor a thermal ones.
Our conclusions are based on numerical results of a real physical
system, so it is a numerical experiment to confirm
the theoretical expectation.

This paper is organized as follows: In Sect.II, we introduce the model. In Sect.III,
we present definitions of several ensembles of states. In Sect.IV, we
obtain our main conclusions from numerical results. In Sect.V and Sect.VI, we show
that our conclusions do not depend on specific lattice configuration and initial
state. In Sect.VII, we study the thermalization of two coupled systems
with different temperatures. In Sect.VIII we study the entanglement properties of the states in thermalization.
Finally a summary of conclusions is presented in Sect.IX.

\section{Model}
Let us first present some fundamental concepts.
Consider a generic quantum many-body system
constituted by weakly coupled system `S' and a large heat bath `B',
in thermodynamic equilibrium,
the state of system is in a canonical form.
In statistical mechanics \cite{Landau}, to obtain this canonical state of `S',
the composite system is described by the microcanonical density matrix
in which each pure state satisfying the constraints of suitable
total energy has equal probability. The state of the
quantum system `S', $\rho ^{(S)}$, can then be obtained by tracing out
the freedoms of the heat bath from the microcanonical ensemble.
By this process,  the quantum state of the system in thermodynamic equilibrium, $\rho ^{(S)}$,
can be found to take
the form of the canonical density matrix which is the thermal state defined as,
\begin{eqnarray}
\Omega ^{(S)}=e^{-\beta \hat {H}^{(S)}}/Z,
\label{canonical}
\end{eqnarray}
where $\beta $ is inverse
temperature, $\hat {H}^{(S)}$ is the Hamiltonian of the system `S',
and the partition function is, $Z={\rm tr}e^{-\beta \hat {H}^{(S)}}$.

\begin{figure}[h]
\includegraphics[height=4.2cm,width=8cm]{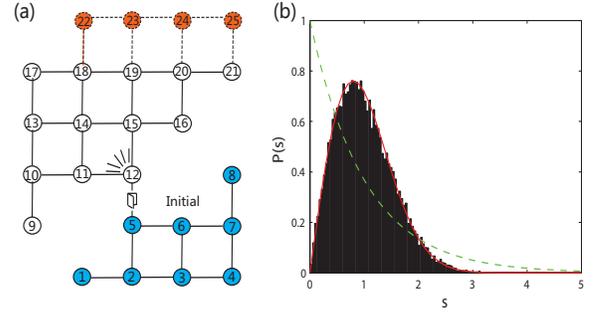}
\caption{(a) Two-dimensional lattice on which five hard-core bosons
propagate in time, the number of lattice sites can be selectively extended from 21 sites
as in Ref.\cite{RigolNature} to 25 sites. (b) Level spacing
distribution of the energy spectrum of the
Hamiltonian (\ref{H}). The superimposed curves show a Poissonian
(dashed green) and Wigner (solid red) distribution, characterizing
integrable and non-integrable (chaotic) systems, respectively.}\label{fig1}
\end{figure}

The Hamiltonian of the system takes the form:
\begin{equation} \label{H}
\hat{H}=-J\sum_{\langle i,j\rangle}(\hat{b}_i^{\dagger}\hat{b}_j+
h.c.)+U\sum_{\langle i,j\rangle}\hat{n}_i\hat{n}_j,
\end{equation}
where $\langle i,j\rangle$ indicates all pairs of nearest-neighbor
sites on, a 21-site, two-dimensional lattice, see Fig.\ref{fig1}(a) for
its configuration,  $J$ is the hopping parameter, $U$ is the
nearest-neighbor repulsion parameter and we set it as, $U=0.1J$.
In Fig\ref{fig1}(a), we consider a portion of the lattice, sites 1 to 8
(lower-right corner) as
the system `S', other parts of the lattice, sites 9 to 21 (upper-left corner)
are considered as the heat
bath `B'. In order to show that our results are general
and do not depend on specific lattice configurations, we also extend
the sites of heat bath `B' from site 21 up to sites 25, see Fig.\ref{fig1}(a).
The coupling between `S' and `B' is through the interaction
between sites 5 and 12, $\hat{H}^{(I)}$, which can be turned on and off.
So Hamiltonian (\ref{H}) can be rewritten as,
\begin{eqnarray}
\hat{H}=
\hat{H}^{(S)}+\hat{H}^{(B)}+\hat{H}^{(I)}.
\end{eqnarray}
As a whole, the coupled system `SB' is an isolated
quantum many-body system. In this paper,
we consider that five hard-core bosons propagate in time on this lattice.

The model (\ref{H}) is in general non-integrable since the
peculiar structure of the lattice in Fig.\ref{fig1}(a) has broken any spatial
symmetry. By diagonalizing the Hamiltonian (\ref{H})
with 21 sites (also up to 25 sites), we find that its energy level spacing
distribution is exactly a Wigner distribution \cite{wigner}, $P_{Wigner}(s)=\frac{\pi}{2}s\exp(-\frac{\pi s^2}{4})$,
see Fig.\ref{fig1}(b), which
corresponds to nonintegrable system (or chaotic system). So thermalization
is generally expected to happen.

\section{Time evolution and ensembles}
We consider the initial state as
$|\psi (0)\rangle $ which is the ground state of
$\hat {H}^{(S)}$ with all five hard-core bosons confined
in eight sites of `S', the interaction $H^{(I)}$ is
in off position initially. We can expand the initial-state
wavefunction by the eigenstate basis of
the whole Hamiltonian $\hat{H}$ as,
$|\psi(0)\rangle=\sum_{\alpha}C_{\alpha}|\Psi_{\alpha}\rangle $,
where $C_{\alpha}=\langle \Psi_{\alpha}| \psi(0)\rangle$ are the overlaps between
the initial pure state and the eigenstates $|\Psi_{\alpha}\rangle$,
the corresponding eigenvalues are $E_{\alpha}$.

Before proceed the dynamical relaxation, we consider different sets of ensembles.
The energy of the system which is conserved in the evolution can be found to be,
$E_0=\langle \psi(0)|\hat {H}|\psi(0)\rangle=-5.06J$. In statistical mechanics with
the equal a priori probability postulate, the
microcanonical ensemble can be defined as,
\begin{equation} \label{micro}
\rho_{\mathrm{micro}}=\frac{1}{{\cal N}}\sum_{\alpha,
|E_0-E_{\alpha}|<\Delta
E}|\Psi_{\alpha}\rangle\langle\Psi_{\alpha}|.
\end{equation}
where ${\cal N}$ is the number of $E_{\alpha}$ in the energy window,
$|E_0-E_{\alpha}|<\Delta E$, and $\Delta E=0.1J$ is relatively small.
The canonical ensemble, which is the thermal state corresponding to $\hat {H}$,
is defined as,
\begin{equation} \label{cano}
\rho_{\mathrm{cano}}=\sum_{\alpha}P_{\alpha}
|\Psi_{\alpha}\rangle\langle\Psi_{\alpha}|,
\end{equation}
with the probability distribution, $P_{\alpha}=e^{-\beta E_{\alpha}}/Z$,
$Z=\sum_{\alpha}e^{-\beta E_{\alpha}}$, $\beta=1/k_B T=1/1.87J$, the
inverse temperature $\beta$ is consistently obtained from the
equation, $\mathrm{tr} (\rho _{\mathrm{cano}}\hat {H})=\sum_{\alpha}e^{-\beta E_{\alpha}}E_{\alpha}/Z
=E_0$. We can also consider the generalized Gibbs matrix \cite{Rigol2007,RigolNature},
\begin{eqnarray}
\rho _{\mathrm{gibbs}}=\sum_{\alpha}|C_{\alpha}|^2|\Psi_{\alpha}\rangle
\langle\Psi_{\alpha}|.
\label{gibbs}
\end{eqnarray}

\begin{figure}[h]
\includegraphics[height=5cm,width=7cm]{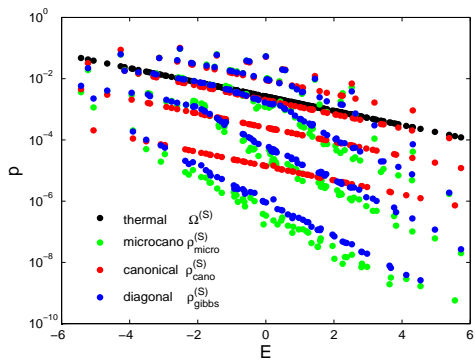}
\caption{Four density matrices of `S' from different ensembles are presented.
With Eq.(\ref{cano}) as an example, the horizontal axes is $E_\alpha $, the
vertical axes is $P_{\alpha }$ in logarithmic scale, the results are almost
linear, so slope corresponds to minus of inverse temperature.}
\label{fig2}
\end{figure}

We will mainly study the state of the quantum system `S'. Corresponding to those ensembles, the density
matrices of `S' can be obtained by tracing out the freedoms of the heat bath `B',
$\rho ^{(S)}_{X}={\mathrm{tr_{B}}}\rho _{X}$, where $X$ denotes the cases of microcanonical,
canonical and Gibbs, respectively. Together with thermal state $\Omega ^{(S)}$ defined
in (\ref{canonical}), these four density
matrices are schematically presented in Fig.\ref{fig2}.
From the statistical mechanics, in thermodynamical limit with infinite large heat bath,
as we mentioned, $\rho ^{(S)}_{\mathrm{micro}}$ should take the form of thermal state $\Omega ^{(S)}$.
However, from Fig.\ref{fig2}, it seems those two states are not like each other.
On the other hand, we may find that
$\rho ^{(S)}_{\mathrm{micro}}$ is more like $\rho ^{(S)}_{\mathrm{gibbs}}$, there
seems no other apparent similarities.
Quantitatively, we can use a Hilbert-Schmidt norm to quantify the distance between
two density operators, $||O||^2\equiv {\rm tr}O^{\dagger }O$, $O$ is the difference
of two matrices. We find,
\begin{eqnarray}
||\rho ^{(S)}_{\mathrm{gibbs}}
-\rho ^{(S)}_{\mathrm{micro}}||^2\approx 10^{-4},
\end{eqnarray}
this confirms our observation
from Figure \ref{fig2},
\begin{eqnarray}
\rho ^{(S)}_{\mathrm{micro}}\approx \rho ^{(S)}_{\mathrm{gibbs}}.
\label{gibbsmicro}
\end{eqnarray}
Other distances are relatively larger ranging from $10^{-3}$ to $10^{-2}$.

Now let's consider the dynamical relaxation. The time evolution begins with a sudden turning on the
the interaction $\hat {H}^{(I)}$ between sites 5 and 12. The initially confined five hard-core bosons
begin to propagate on the whole lattice. The
wavefunction then evolves, $|\psi(t)\rangle=e^{-i\hat{H}t}|\psi(0)\rangle $, we have,
\begin{eqnarray}
|\psi(t)\rangle=\sum_{\alpha}C_{\alpha}e^{-iE_{\alpha}t}
|\Psi_{\alpha}\rangle .
\label{evolve}
\end{eqnarray}
Our numerical experiment is to study the dynamical properties of
wavefunction $|\psi (t)\rangle $, in particular the time dependent
reduced density operator of system `S' compared with that of
the states from various ensembles. In thermodynamic equilibrium,
we can also compare our result with that of the general
canonical principle \cite{winter}.

\section{Thermalization and general canonical principle}
\begin{figure}[t]
\includegraphics[height=5cm,width=8cm]{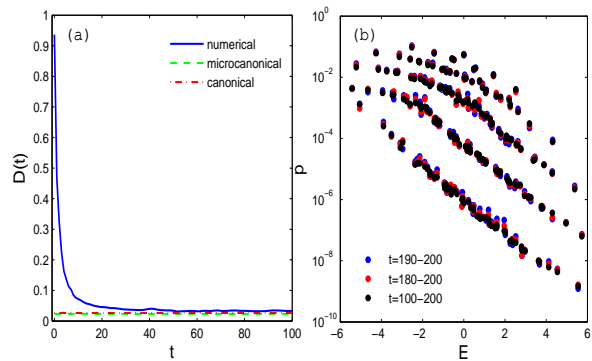}
\caption{(a) The time evolution of trace norm $D(t)$ between $\rho ^{(S)}(t)$ and thermal
state $\Omega ^{(S)}$. The distances between thermal state with that
of microcanonical and canonical ensembles are presented for comparison.
(b) Time average of reduced density matrices $\overline{\rho ^{(S)}(t)}$ for
different scales of time period. All density matrices are almost the same.}
\label{norm}
\end{figure}

As mentioned in introduction,
thermalization means that any pure states will relax to pure states in which the values
of macroscopic quantities are stationary and universal.
By general canonical principle \cite{winter}, it means that in thermodynamic equilibrium,
from instantaneous pure state (wavefunction) $|\psi (t)\rangle $,
the state of `S', $\rho ^{(S)}(t)$ equals approximately to thermal state $\Omega ^{(S)}$
defined in (\ref{canonical}).
Numerically, we plot in Fig.\ref{norm}(a) the time evolution of distance between two matrices,
$D(t)=||\rho ^{(S)}(t)-\Omega ^{(S)}||^2$. For comparison,
we also present the distances of $\rho ^{(S)}_{micro}$ and $\rho ^{(S)}_{cano}$
with $\Omega ^{(S)}$. As expected, the time dependent distance quickly drops to almost
zero. We confirm this result with different measures of distance, the same
conclusion can be made. Thus by thermalization, roughly,
we confirm the general canonical principle by that any
instantaneous pure state of `SB' after thermalization leads approximately to the thermal state of `S'.

However, the following two important questions need be analyzed carefully by our
numerical experiment:(i) Which state should be our goal state for comparing in time evolution,
thermal state $\Omega ^{(S)}$ or $\rho _{\rm micro}^{(S)}$? (ii) What is the state of time average?
Also we need to check what is the case for different initial states.
As we already found that $\rho _{\rm micro}^{(S)}$ is different from thermal state
$\Omega ^{(S)}$ in our system. Of course, it can be expected that those two states
are the same if the freedoms of the heat bath `B' are large enough.
However for our finite two-dimensional lattice, the 13-site heat bath `B' is not
large enough compared with 8-site system `S', so the first question arises.
In the proof of general canonical principle in Ref.\cite{winter}, both ensemble
and time average is not necessary, only the distance between reduced density
operator of a typical pure state and the thermal state is used.
Then in our system, what is the role of
time average, thus the second question arises. We will clarify those questions
by numerical results.

\begin{figure}[t]
\includegraphics[height=4cm,width=8cm]{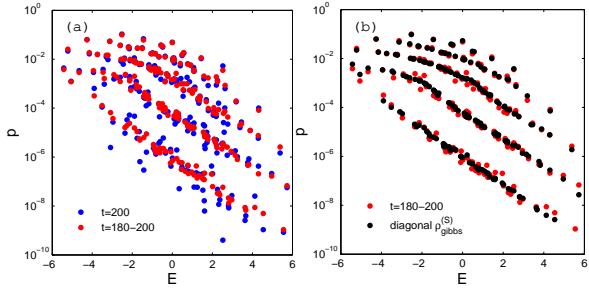}
\caption{(a) The time average density matrix agrees with the instantaneous density matrix.
(b) The time average density matrix agrees well with the density matrix from
generalized Gibbs ensemble.}
\label{avinst}
\end{figure}

From the time evolution of distance in Fig.\ref{norm}(a), we can guarantee that
thermalization happens and the state is stationary after $t=100$.
Obviously, we can find
that the density matrices of time average for different scales of
time period are almost the same, see Fig.\ref{norm}(b). Then it is reasonable to
expect that after thermalization, the time average density matrix is approximately
equal to the instantaneous
density matrix from a single pure state,
\begin{eqnarray}
\overline{\rho ^{(S)}(t)}\approx \rho ^{(S)}(t).
\end{eqnarray}
From the view point of distance, this can be confirmed in Fig.\ref{norm}(a).
The comparison between time average density matrix with the
instantaneous reduced density matrix is, to some extent, obvious and agrees
well this result, see Fig.\ref{avinst}(a).

With the time evolution wavefunction (\ref{evolve}), the density matrix is simply,
$\rho(t)=\sum_{\alpha,\beta}
C_{\alpha}^{*} C_{\beta} e^{i(E_{\alpha}-E_{\beta})t}
|\Psi_{\alpha}\rangle \langle\Psi_{\beta}| $.
Considering thermalization happens which possesses a stationary state,
a long-time average of density matrix of `SB' is expected to be,
\begin{equation} \label{rhot}
\overline{\rho (t)}=\sum_{\alpha}|C_{\alpha}|^2|\Psi_{\alpha}\rangle
\langle\Psi_{\alpha}|.
\end{equation}
This is the generalized Gibbs ensemble (\ref{gibbs}).
And the equivalence is confirmed
by directly comparing those two density matrices, see Fig. \ref{avinst}(b).

Remarkably, bear in mind that microcanonical ensemble is the
same as the generalized Gibbs ensemble as shown in (\ref{gibbsmicro}),
we now reach a chain of equations from our numerical
experiment step by step,
\begin{eqnarray}
\rho ^{(S)}(t)\approx \overline{\rho ^{(S)}(t)}\approx
\rho ^{(S)}_{\mathrm{gibbs}}
\approx \rho ^{(S)}_{\mathrm{micro}}.
\end{eqnarray}
Here we emphasize that this equation chain is not only based on
the measure of distances, but rather based on the exact form of those
density matrices which is, of course, more explicit and more precise.
While for distance as in Fig.\ref{norm}(a), we can not even distinguish
the cases of canonical ensemble and the microcanonical ensemble.

As in Ref.\cite{winter}, next we should follow the standard statistical mechanics
to obtain the thermal state from the state of microcanonical ensemble.
The difference between $\rho ^{(S)}_{\rm micro}$ and thermal state $\Omega ^{(S)}$
should be a finite-size effect for our system.
In this paper, we show that
the equal a priori probability postulate can be replaced
by a numerical experiment which confirms the general canonical principle.
In addition, dynamically, our result is from a process of thermalization.
We show any pure states relax to pure states corresponding to microcanonical
ensemble with stationary and universal macroscopic values. To show the
results are independent of a wide range of initial states, we next will set
the initial state differently,  we will show that
the conclusion still holds.
Also we can consider different lattice
configurations.

\section{Thermalization with different lattice configurations}
\begin{figure}[h]
\includegraphics[height=5cm,width=7.5cm]{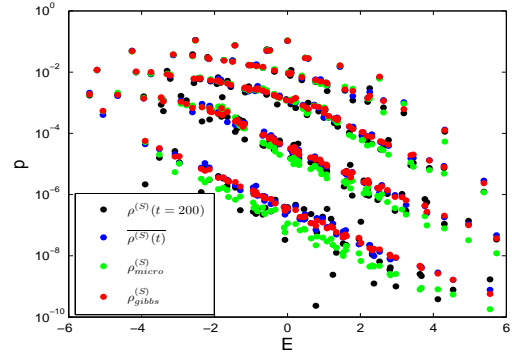}
\caption{Results with 25-site lattice. Four density matrices of `S' from instantaneous pure state after thermalization,
time average after thermalization, the generalized Gibbs ensemble and the microcanonical
ensemble are presented. They almost overlap with each other which confirms our
conclusion.}
\label{supfig1}
\end{figure}

We expect that our results do not depend on specific lattice configurations.
So we cast the heat bath `B' with changeable number of freedoms, i.e. the
sites 22-25 can be selectively added as part of the heat bath `B', see Fig\ref{fig1}(a).
Our main conclusions expressed as a chain of equations always hold,
\begin{eqnarray}
\rho ^{(S)}(t)\approx \overline{\rho ^{(S)}(t)}\approx
\rho ^{(S)}_{\mathrm{gibbs}}
\approx \rho ^{(S)}_{\mathrm{micro}}.
\label{supchain}
\end{eqnarray}
Fig.(\ref{supfig1}) is for case of 25-site lattice, all results are similar
for cases with 21, 23, 24, 25 sites, respectively.

\begin{figure}[h]
\includegraphics[height=5cm,width=7.5cm]{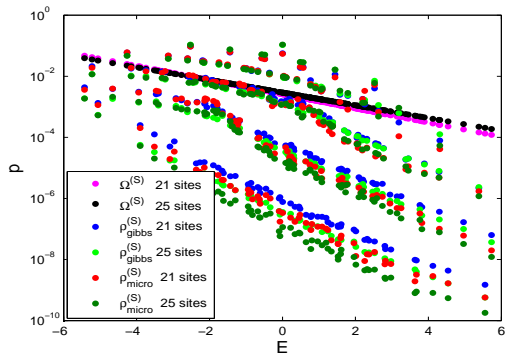}
\caption{Results for 21-site lattice and 25-site lattice.
Three density matrices of `S' are presented:
time average  after thermalization, the generalized Gibbs ensemble and the microcanonical
ensemble.}
\label{supfig2}
\end{figure}

Next we present the states of `S' for different lattice configurations, in particular, with
number of lattice site of heat bath `B' changing, see Figure (\ref{supfig2}). Still thermal state
and the case of microcanonical ensemble themselves are different, it seems that our
system is not large enough. In principle, this difference is due to finite-size effect,
in particular, our system is a general physical system with thermalization,
the interaction between `S' and `B' is weak. So thermodynamical limit can be
obtained when heat bath takes the limit to have infinite degrees of freedom.

\begin{figure}[h]
\includegraphics[height=10cm,width=8cm]{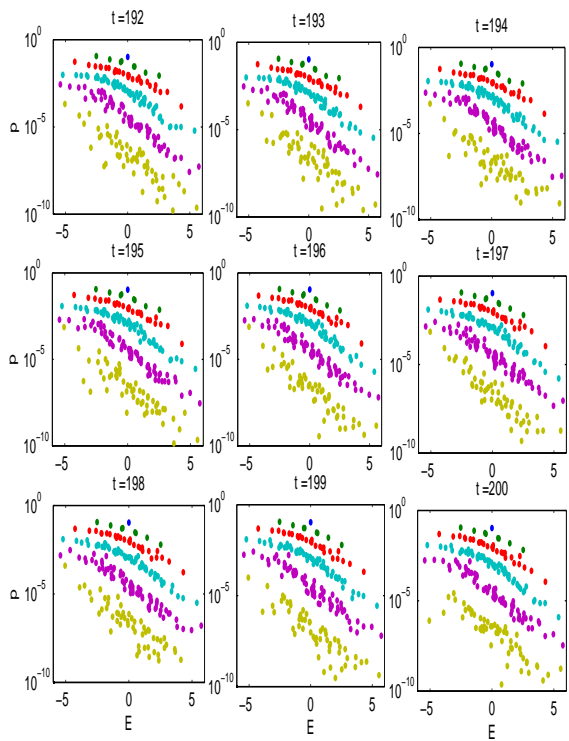}
\caption{Results of 25-site lattice: density matrices of `S' from
instantaneous pure state at different time.}
\label{supfig3}
\end{figure}

The state of `S' from the instantaneous pure state is almost stable and is approximately like from the
the generalized Gibbs ensemble, the quantum fluctuation
is almost negligible. One can find in Figure (\ref{supfig3}),
states from instantaneous pure state in different time after thermalization are almost the same.

So by those results, we find that our conclusions do not depend on specific lattice configurations,
after thermalization,
the state of `S' is stable with negligible quantum fluctuation.

\section{Thermalization with different initial states}
Our conclusion is that from any instantaneous pure state after thermalization, the
state of system `S' is like from a microcanonical ensemble. By definition, we know
that the microcanonical ensemble depends only on the energy of the initial state with
a narrow energy window. So, we should expect that a wide range of initial states
with fixed energy should relax to pure states with similar properties as the microcanonical
ensemble.  On the other hand, the state of time average is equivalent with
the generalized Gibbs ensemble which depends on the distribution of $C_{\alpha }$ of
eigenstates $|\Psi _{\alpha }\rangle $ for the initial state. It is thus initial state
dependent. So it is necessary for numerical experiment to confirm our conclusion for
different initial states.

We consider the following six different initial states with approximately equal energy $E_0$.
Still five hard-core bosons are on the lattice, however, initially the number of hard-core bosons
in `S' are: 5, 4, 3, 2, 1 and 0, and other hard-core bosons are confined in
lattice sites of heat bath `B', respectively. There is no interaction between `S' and `B'
before time evolution, $\hat {H}^{(I)}$ , and we set the initial states as product (separable) pure states,
$|\psi ^{(S)}(0)\rangle \otimes |\psi ^{(B)}(0)\rangle $. The energies of `S' and `B' are,
$\langle \psi ^{(S)}(0)| \hat {H}^{(S)}|\psi ^{(S)}(0)\rangle \equiv E_0^{(S)}$, and
$\langle \psi ^{(B)}(0)| \hat {H}^{(B)}|\psi ^{(B)}(0)\rangle \equiv E_0^{(B)}$, respectively.
However, the total energy is $\langle \psi ^{(S)}(0)|\otimes \langle \psi ^{(B)}(0)|
 \hat {H}|\psi ^{(S)}(0)\rangle \otimes |\psi ^{(B)}(0)\rangle \equiv E\approx E_0$, which
 is slightly different from the summation of energies of system `S' and heat bath `B'.
 This is because of the interaction $\hat {H}^{(I)}$, which should be nonzero but must be weak in the
 thermalization.  Explicitly, for our finite-size system,
we have,
\vskip 1truecm
\begin{tabular}{|c|c|c|c|c|}
  \hline
  $S$ &$E_0^{(S)}$& $E_0^{(B)}$ &$E_0^{(S)}$+$E_0^{(B)}$ & $E$ \\
  \hline
  5 & -5.0595 & 0 & -5.0595 & -5.0595 \\
  \hline
  4 & -5.4262 & 0.3843 & -5.0419&-5.0406  \\
  \hline
  3 & -5.2228 & 0.1943 & -5.0286 & -5.0202\\
  \hline
  2 & -4.2627 & -0.8044 & -5.0672 & -5.0619 \\
  \hline
  1 & -2.5231 & -2.5355 & -5.0586 & -5.0560 \\
  \hline
  0 & 0 & -5.0692 & -5.0692 & -5.0691 \\
  \hline
\end{tabular},
\vskip 1truecm
\noindent where $S$ in table is the number of bosons initially in `S'. We find the
interaction between system `S' and heat bath `B' is weak, and so we have,
\begin{eqnarray}
E_0\approx E\approx E_0^{(S)}+E_0^{(B)}.
\end{eqnarray}

\begin{figure}[h]
\includegraphics[height=11cm,width=8cm]{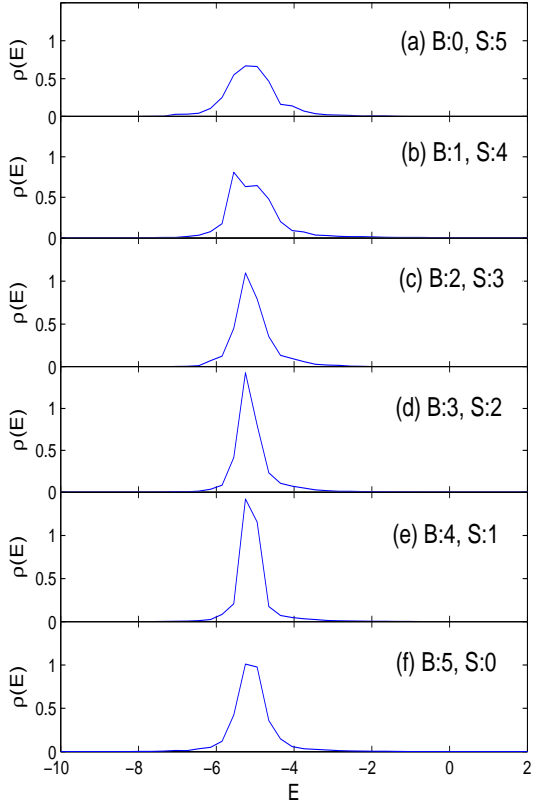}
\caption{The distributions of $C_{\alpha }$ for different initial states, where initially the number
of hard-core bosons on `S' are 5,4,3,2,1 and 0, respectively}
\label{supfig4}
\end{figure}

\begin{figure}[ht]
\includegraphics[height=5cm,width=7cm]{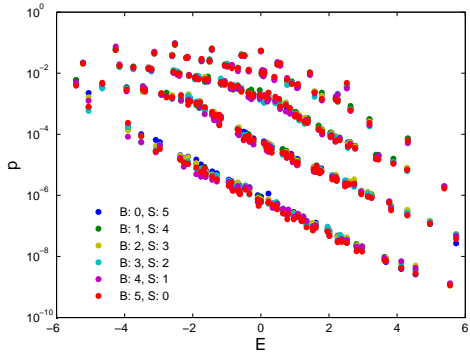}
\caption{The generalized Gibbs ensembles are almost the same for different initial states,
the number of hard-core bosons on `S' are 5,4,3,2,1 and 0, respectively.}
\label{supfig5}
\end{figure}

The generalized Gibbs ensemble is defined by the distribution of $C_{\alpha }$
which is the overlap between initial state and the eigenvectors $|\Psi _{\alpha }\rangle $ of the
total Hamiltonian. Figure (\ref{supfig4}) shows the distribution of $C_{\alpha }$ for different
initial states, so the generalized Gibbs ensembles for different initial states are
almost the same, see Figure (\ref{supfig5}).

\begin{figure}[ht]
\includegraphics[height=11cm,width=8cm]{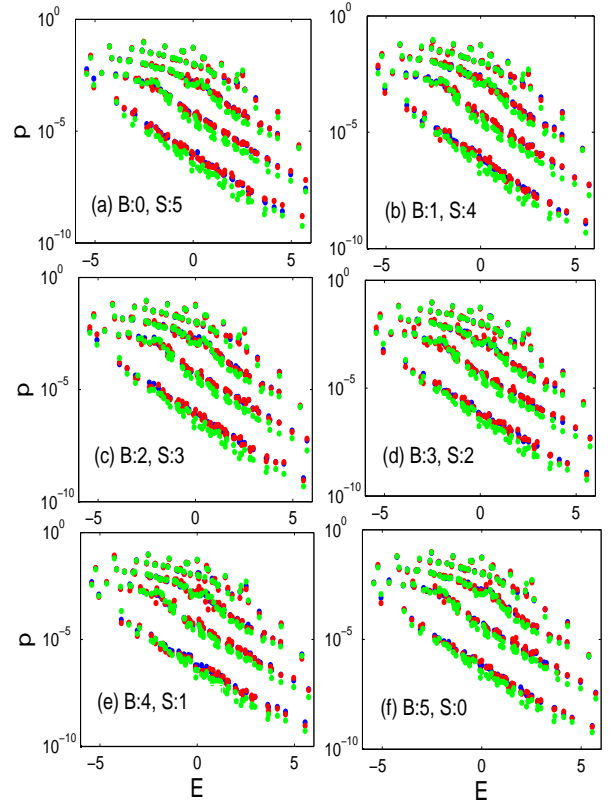}
\caption{Three density matrices of `S', which are
time average after thermalization, the generalized Gibbs ensemble and the microcanonical
ensemble, are presented. They agree well with each.}
\label{supfig6}
\end{figure}

To analyze our numerical data, we should confirm that our conclusion
for different initial states should always hold.
The numerical results are presented in Figure (\ref{supfig6}), here the time average
is for short period of time so that it is approximately the instantaneous pure state.
Since the microcanonical ensemble with fixed energy is independent of the specific
initial state, the main point need be checked carefully is whether the case of microcanonical
overlaps with the case of generalized Gibbs ensemble or the case of time average.
We find that they all are almost the same. Thus we conclude that
for a wide range of initially states, the state of `S' is always like from
a microcanonical ensemble.

\begin{figure}[ht]
\includegraphics[height=5cm,width=7cm]{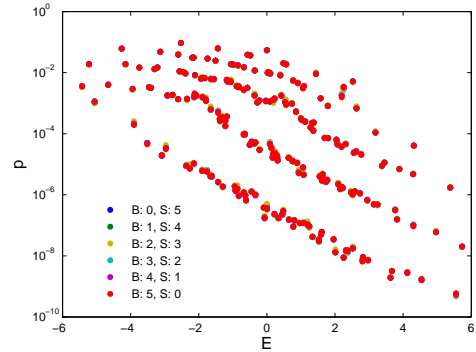}
\caption{With energies being almost the same, states from the microcanonical
ensembles overlap with each other. }
\label{supfig7}
\end{figure}

Figure (\ref{supfig7}) shows that with energy fixed, the microcanonical ensembles are
almost the same, this is in accordance with the definition.

\section{thermalization of two coupled systems with different temperatures}
In order to study the thermalization from different points of view, we next consider
that the coupled systems `S' and `B' initially are in different temperatures and the
corresponding initial states are mixed states, this is in comparison that
of previous sections, where the state of the whole system `SB' is always pure.
Since thermalization can happen
in this coupled system, a dynamical equilibrium can be reached and the whole system `SB'
should have a same temperature.
Note that
thermalization is closely related with quench where the Hamiltonian of the
system changes quickly and the state of the system then evolves under the
new Hamiltonian.

In numerical calculations, we assume that, $\beta ^{(S)}=1$, $\beta ^{(B)}=2$,
which are inverse temperatures of systems `S' and `B', respectively. We still restrict
us to the case of five hard-core bosons, two of them are confined initially
in `S' and the remained three are in `B' system. The initial state of `S' is the thermal
state as presented in (\ref{canonical}) with $\beta =1$,
\begin{eqnarray}
{\Omega '}^{(S)}=e^{-\hat {H}^{(S)}},
\end{eqnarray}
where the normalization of partition function is omitted without confusion.
Similarly initial state of `B' is also a thermal state,
\begin{eqnarray}
{\Omega '}^{(B)}=e^{-2\hat {H}^{(B)}},
\end{eqnarray}
where $\beta ^{(B)}=2$. It is clear that both states are mixed. The thermalization begins
with the turning on the interaction between systems `S' and `B' by term $\hat {H}^{(I)}$.

\begin{figure}[ht]
\includegraphics[height=6cm,width=8cm]{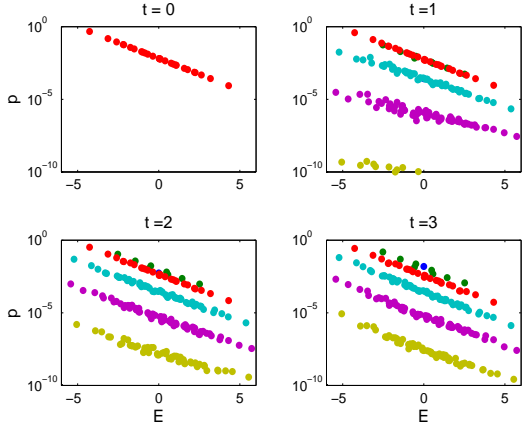}
\caption{State of the system `S' at time points $t=0,1,2,3$ starting from a thermal state.}
\label{mixed1}
\end{figure}

\begin{figure}[ht]
\includegraphics[height=6cm,width=8cm]{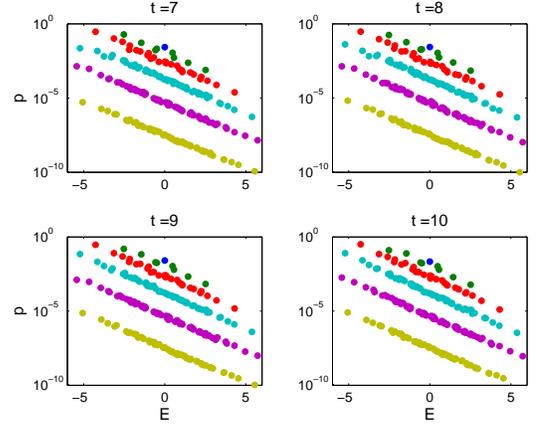}
\caption{State of the system `S' at time points $t=7,8,9,10$ starting from a thermal state.
Thermalization happens and the instantaneous states of `S' are almost the same which shows
the equilibrium reaches.}
\label{mixed2}
\end{figure}

Figure (\ref{mixed1}) shows that the state of `S' is initially a thermal state
at time $t=0$, then thermalization happens, the state changes at points $t=1,2,3$.
The dynamical equilibrium quickly reaches, we find that the state of `S' at
$t=7,8,9,10$ is almost stable, which is shown in Figure (\ref{mixed2}).

\begin{figure}[ht]
\includegraphics[height=5cm,width=7cm]{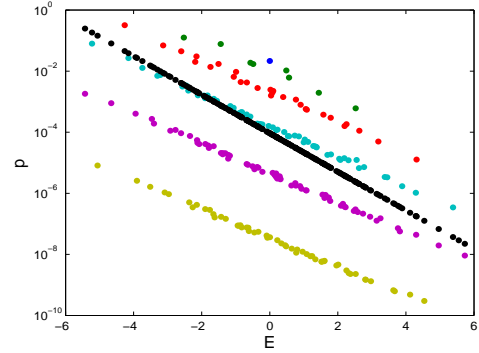}
\caption{State of the system `S' at time points $t=10$ and the reduced density operator
of `S' (black) from a thermal state with $\beta =1.4540$.}
\label{mixed3}
\end{figure}

If we study the whole system `SB', corresponding to different temperatures of `S' and `B'
systems, the average temperature can be calculated as $\beta =1.4540$. After thermalization,
the state of `S' is still different from a thermal state. Figure (\ref{mixed3}) shows that
the state of `S' at time $t=10$ is different from a thermal state though it starts from
a thermal state. Here we argue that this is a finite-size effect, in case the heat bath
`B' is large enough, it will be like a thermal state.

The involved states in this section are mixed states. Here we show that a dynamical equilibrium
can be reached and thermalization can happen under different conditions.

\section{Thermalization and decoherence}
\begin{figure}[t]
\includegraphics[height=6cm,width=9cm]{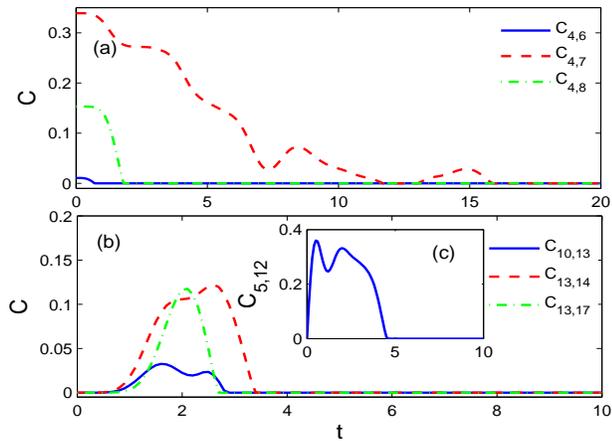}
\caption{The time evolution of entanglement measured by concurrence $C$ for pairs of sites.
For example, $C_{4,6}$ is the concurrence between sites 4 and 6.}\label{con1}
\end{figure}

In a generic isolated quantum system,
thermalization of state `S' which is
initially a pure state, is induced by the interaction between the system `S' and
the heat bath `B' which will destroy the coherence of the initial state.
Thus entanglement in the initial pure state, in general, will decrease
since of the decoherence from the heat bath. Still many aspects of entanglement
are of interest in the time evolution in `S' and `B'.
We use concurrence, one measure of entanglement \cite{wootters},
to quantify the entanglement between
different pairs of sites on the lattice, each is a qubit-qubit state.
The results are shown in Fig.\ref{con1}. Here we would like to point
out several interesting facts: (a) As expected, entanglement
between nearest-neighbor sites, $C_{4,7}$, is relatively larger,
while in time evolution, it does not decrease monotonically which indicates
a non-Markovian dynamics. (b) With thermalization, all pair-type entanglement
converge to almost zero including the sites in heat bath `B' and the coupling
states of (5,12), though the whole system is always
a 21-partite pure state. This may indicate a property of thermal state which
can be described classically so without entanglement.
(c) The entanglement of pairs in `B', is relatively
large for a period in the beginning of thermalization while decrease finally to zero.
This indicates the quantum aspects of the thermalization for an isolated quantum
system, it also indicates that entanglement may be necessary in the process to relax
to a stationary state.

Besides pair-wise entanglement, it is also of interest to study the local
von Neumann entropy as measure of entanglement between coupled systems.
Since initially the state of `S' is a pure state, when the interaction
between `S' and `B' is turned on, the von Neumann entropy of `S' increases quickly and
reaches almost a constant. This is also an indication of thermalization, i.e. thermalization
can be measured by chaotic of system `S' quantified by local von Neumann entropy. 
We remark that since the overall
state is pure, this von Neumann entropy is a measure of entanglement between systems `S' and `B'.

\section{Conclusions and discussions}
In conclusion, by numerical experiment,
we prove a chain of equations and reach a conclusion that after thermalization,
the instantaneous density operator of `S' is equivalent to the state
from a microcanonical ensemble. Thus dynamically, we prove the
general canonical principle which can replace
the equal a priori probability postulate for microcanonical ensemble
in statistical mechanics. We use the density matrix itself rather than
distance to analyze our results thus their derivations are
more explicit and precise, we identify that the difference
between state from the microcanonical ensemble and the thermal state
is a finite-size effect.
In the generic isolated quantum system,
we systematically present various ensembles and clarify their relations
numerically. In addition, we show in the thermalization, though
the state of `SB' is always a pure state, its pair-type entanglement
of system `S' is almost zero like a classical state, but it jumps from zero
in the beginning of thermalization to finite value and drops again
to zero when thermal equilibrium is finally reached.
This shows that entanglement may be useful for thermalization.

The main method in this paper is to compare density matrices schematically,
however, the result agrees well with method of calculating the distances
evaluated by such as Hilbert-Schmidt norm as used in Ref.\cite{winter}.
It is already shown in Ref.\cite{RigolNature} that the one-dimension momentum
distributions are close to that of microcanonical ensemble. 
Since expectation values of observables are based
on measurement operators applied on density matrix,
the method by comparing directly density matrices seems more complete and explicit.

The main conclusion of this paper is based on a real physical system in which
thermalization happens. The process of thermalization is realized by a
numerical experiment so that the result does not depend on any
assumptions. Considering that the system is generic, our conclusion
should be general.

We would like to thank useful discussions with C. P. Sun and H. Dong.
This work is supported by NSFC and ``973'' program (2010CB922904, 2011CB921500).

\end{document}